\documentstyle[12pt]{article}
\newcommand{\be}{ \begin{eqnarray}}
\newcommand{\ee}{\end{eqnarray}}
\newcommand{\beno}{ \begin{eqnarray*}}
\newcommand{ \eeno}{\end{eqnarray*}}
\newcommand{\raf}[1]{(\ref{#1})}
\begin{document}
\bibliographystyle{try}
 \begin{titlepage}
\hspace{11cm}
{\large SUNY-NTG 94-26}
\vspace{.7cm}
\begin{center}
\ \\
{\large {\bf Cold Kaons from Hot Fireballs}}
\vspace{2cm}
\ \\
{\large Volker Koch}
\ \\
\ \\
{\it Physics Department, State University of New York\\
Stony Brook, NY 11794, U.S.A.}\\
\ \\
\today \\

\vspace{2cm}
{\large \bf Abstract}\\
\vspace{0.2cm}
\end{center}
 \begin{quotation}
The E814-collaboration has found a component of very low $m_t$ $K^+$ mesons
with a slope parameter of $T \sim 15 \, \rm MeV$. We
will present a scenario which explains the observed slope parameter
and which allows us to predict the expected slope parameter for
 kaons produced in heavier systems such as Au+Au. Implications for the
restoration of chiral symmetry in relativistic heavy ion collisions are
discussed.
 \end{quotation}
\end{titlepage}
\newpage
\section{Introduction}
\indent
At quark matter '93 J. Stachel \cite{Sta94}
presented the first data on kaon spectra
obtained by the E-814 spectrometer at the AGS. These data cover only the very
lowest range in $m_t$ ($m_t - m_k \leq 25 \, \rm MeV $) at somewhat forward
rapidities ($ 2.2 \leq Y \leq 2.6$) and they can be parameterized
by an exponential in $m_t$ with a slope parameter as low as $T \simeq  15 \,
 \rm MeV$. This slope parameter should be compared with the finding of the
E-802 collaboration \cite{Abb90} which  gives a slope-parameter of $\sim 150
\, \rm MeV$ at central rapidities for values of $m_t - m > 100 \, \rm MeV$.
Also p-Be data, which are essentially equivalent to p-p,
show a slope parameter of the order of $150 \, \rm MeV$ \cite{Abb91}.
Hence, the
slope parameter
found by E-814 certainly has to have its origin ion the many body dynamics of
the heavy-ion collision.

A steep rise in the spectrum at low transverse momenta has also been observed
for pions \cite{Abb90,Ahm92,Hem94} .
In this case the so called low $p_t$-enhancement
can be understood by the decay of the delta resonance and event generators
which include the deltas such as RQMD and ARC \cite{SSG89,SKP92}
do reproduce the
observed spectra very well. In case of the kaons $(K^+$), however, a component
due to resonance decay is essentially ruled out, since there very few Phi
mesons and strange anti-baryons are expected to be produced.

Another mechanism, which may lead to a cold component in the spectrum is the
effect of attractive potentials as has been discussed in several papers
\cite{KBK91,Shu91,XKK93,FKB93,KB93a}.
In references \cite{KBK91} and \cite{FKB93} it
was shown that the presence of an attractive mean field can affect the slope
parameter of the particle spectra, in particular the high $p_t$ part. If,
however, the mean field is to affect the soft part of the spectrum,
as suggested by ref. \cite{Shu91} for the soft pion spectrum at
CERN - energy heavy ion collisions, a subtle interplay of the expansion time
of the
fireball and the particles under consideration is needed, as discussed in
detail
in ref. \cite{KB93a}. This relation essentially is
the requirement for adiabaticity of the expansion in thermodynamics, meaning
the expansion velocity has to be small compared to the velocity of the
particles to be cooled. While at CERN  energies the requirement for
adiabaticity is not fulfilled, as demonstrated in ref. \cite{KB93a}, at lower
energies such as the Bevalac, these conditions hold and a cooling due to the
mean field is possible \cite{XKK93}.

In this note we will discuss under which circumstances an adiabatic cooling of
the $K^+$ is possible. We will argue, that the observed very cold component in
the kaon spectrum requires a rather small expansion velocity of the fireball,
which, at the observed temperatures, can only be achieved if one assumes that
the system initially goes through a phase with high energy density but low
pressure. Such a phase is to be expected from the chiral restoration
transition. As discussed in a recent analysis of lattice gauge data
\cite{KB93b}
the chiral restoration transition involves the decondensation of the gluon
condensate which gives rise to an effective bag pressure of $p_{bag}  \simeq
- 200 \, \rm MeV/fm^3$.
This bag- or decondensation pressure lowers the pressure of the
system to very small values, nearly zero, and certainly much smaller than the
$1/3$ of the energy density for an ideal gas.
As a
consequence the system expands much more slowly than an ideal gas, which,
for temperature of the order of $150 \, \rm MeV$,
has an expansion velocity of the order of $1/3$ of the velocity of light.
At this point we should also mention that the lowering of the pressure due to
decondensation is inherently absent in cascade-type calculations. Those
essentially simulate the behavior of an ideal gas and thus,
as we will show below,
predict too large an expansion velocity in order for adiabatic cooling to be
effective.

Another ingredient of our model is an attractive potential for the $K^+$.
Since the $K^+N$ scattering amplitude is
repulsive simple impulse approximation predicts a slightly
repulsive optical potential in nuclear matter at zero temperature
\cite{LSW93,LJM94}
\be
U = \frac{- 4 \pi }{2 m_K} (1 + \frac{m_k}{M_N} )\, a_{K^+N} \,  \rho
\sim 25 MeV \frac{\rho}{\rho_0}
\label{eq.1.1}
\ee
where $a_{K^+N} \simeq -0.2 \, \rm fm$ is the isospin averaged s-wave
kaon nucleon scattering length \cite{DW82}. In the framework of chiral
perturbation theory \cite{NK87},
this potential arises from the interplay of an attractive
scalar piece, which is due to explicit chiral symmetry breaking,
and a repulsive
vector potential as a result of vector meson exchange. If for some reason at
high temperature the vector meson coupling is reduced, the resulting kaon
potential could very well turn out to be attractive. Actually lattice gauge
results on the behavior of the baryon number susceptibility can be understood
in a hadronic framework only if the vector mesons decouple above $T_c$
\cite{Kun91}. (For a more detailed discussion see \cite{BR94}.)
It is, however, not the purpose of this article
to develop yet another theory about the
behavior of the $K^+$ in matter. We rather want to follow a phenomenological
approach and determine the strength of the $K^+$ optical potential from the
available data.  As we will
demonstrate in the following section, this can be done rather unambiguously
within the framework of our model of collective cooling. Our result for the
kaon potential should, therefore, be viewed as a phenomenological one, which
then needs to be understood from underlying theoretical principles.

This article is organized as follows. In the following section we will present
a
schematic model of the
expansion, which explains the essential physics of cooling the kaons
and shows that the model parameters can be fixed from existing data. In section
3 we present results of a more microscopic calculation using relativistic
transport theory \cite{BKM93} and we also present out prediction for the larger
system $Au + Au$ after having fixed the model parameters by comparing with
existing data on $Si+Au$.

\section{A schematic model}

It is widely accepted that in AGS collisions a hot zone consisting
mainly of nucleons, deltas, and pions is being formed.  In our
schematic model we assume that these particles, in particular the nucleons and
deltas, give rise to an attractive potential for the kaons
which follows the local baryon density
\be
U(\rho(r)) \sim \rho(r)
\label{eq.2.1}
\ee
where $\rho(r)$ is the local baryon density. As a result of the expansion of
the
fireball, the baryon-density and hence the potential change with time.
We assume a Woods-Saxon type density profile.
\be
\rho(r) \sim  \frac{1}{\exp((r-r_0(t))/\Delta(t) ) + 1}
\label{eq.2.2}
\ee
so that the potential has the following form
\be
U(r) =  \frac{U_0}{\exp((r-r_0(t))/\Delta(t) ) + 1}
\label{eq.2.2b}
\ee
The expansion of the fireball is
then modeled by increasing the radius $r_0$ and the surface thickness $\Delta$
as a function of time.
\be
r_0(t) &=& r_0(t=0) + \frac{t}{\tau}
\nonumber \\
\Delta(t) &=& \Delta(t=0) \frac{r_0(t)}{r_0(t=0)}
\nonumber \\
U_0(t) &=& U_0(t=0) \left( \frac{r_0(t=0)}{r_0(t)} \right)^3
\label{eq.2.3}
\ee
where $v=1/\tau$ is the expansion velocity.
By scaling the surface thickness and the central value of the potential with
the radius we ensure that the integral over
the density distribution, i.e. the baryon number remains constant with time.
For the Si+Au case we assume the following initial conditions for the fireball

\be
R = 3.5 \, \rm fm
\nonumber \\
\Delta = 1 fm
\label{eq.2.4}
\ee
The potential and its time dependence is then determined from eq. \raf{eq.2.1}
using the above density
function.

While the motion of the baryons and the resulting mean field potential for the
kaons is modeled by the time dependence of the density distribution,  the
propagation of the kaons in the potential is treated explicitly by solving the
appropriate semiclassical Vlasov equation.
As it is common practice in the BUU/VUU type (see. e.g.
\cite{BKM93}, \cite{BD88})
transport theoretical description, the Vlasov equation is solved by the
so called test particle method, in which the phase space distribution is
represented by those test particles. The solution of the Vlasov equation is
then equivalent to solving
classical Newton equations of motion for these test particles in the potential
given above.
The initial  distribution of these kaon test particles follows
the same density profile \raf{eq.2.2} in
coordinate space. In  momentum space the kaons are distributed according
to a thermal Bose-distribution with temperature
\be
T = 170 \,\rm MeV
\label{eq.2.6}
\ee
and zero chemical potential. In order to obtain reasonable statistics for the
spectra at least 50000 test particles are needed. Since we are only interested
in the shape of the spectra and not the absolute yield, a normalization of the
phase space distribution is not required.

Given the above initial condition, we have two parameters left to play with,
the expansion time $\tau$ and the depth of the potential $U_0$. As we will
show,
 both these parameters can be fixed by comparison with the data.
To this end,
let us discuss their effect on the spectra. In fig. \ref{fig.2.1} we show the
initial (full histogram) and the final (after expansion) spectrum at fixed
potential
depth of  $U_0 = -50 \, \rm MeV$ and for three different
expansion times, $\tau = 10 fm/c$, $\tau = 5 fm/c$,
and $\tau = 2.5 fm/c$.
We find, that the slope at low transverse mass strongly depends on the
expansion time used: the longer the expansion time the smaller the slope
parameter, exactly what one expects from adiabatic expansion. The depth of
the potential, on the other hand, hardly affects the slopes at low
transverse mass but rather controls the number of particles contributing to the
soft component of the spectrum or, equivalently,
the value of the transverse mass
where the spectrum changes to the initial slope. The reason why the depth of
the
potential hardly affects the slope at low transverse masses can be understood
classically. Particles are being
slowed down only by the gradient of the potential,
the force, which is effective only at the surface. If the system expands too
quickly, the surface moves away from the kaons, and thus cannot slow them down.
If the expansion is slow, on the other hand, only those kaons
which are trapped inside the potential, will be cooled adiabatically.
The number of particles
`trapped' inside the potential naturally depends on the depth of the potential
as only those with a kinetic energy smaller than the potential
will be kept inside.
Therefore, the number of particles contributing
to the soft component of the spectrum is directly related to the depth of the
potential. This is demonstrated in fig. \ref{fig.2.2}, where we show the
resulting spectra for the different potentials $U_0 = -50 \,\rm MeV$,
$U_0 = -100 \,\rm MeV$,
and $U_0 = -150 \,\rm MeV$. We see that the choice of potential hardly affect
the slope of the soft component.

{}From the above consideration it is clear, that both
model parameters can be fixed from an experimental spectrum
with a very low temperature
component at small transverse momenta and a high temperature component at
large transverse momenta. The expansion
time is controlled
by the slope of the low temperature component and the depth of the
potential by the position at which the spectrum changes from cold to hot.
Our schematic model, of course, is not realistic enough, to compare with data.
To
this end, we will use a relativistic transport description \cite{BKM93} which
takes into account the rescattering of the kaons in the fireball
as well as the proper transport properties of the particles
(nucleon, deltas and pions) which make up the fireball. In
addition, the transport description makes sure, that the expansion of the
fireball conserves energy and momentum, which has not been addressed in our
schematic consideration.

\section{More realistic model}
In this section we want to discuss the transport theoretical calculation of
the expansion and present the results for the kaon spectra. As already
indicated in the introduction, ideal gas type expansion is to fast to be
effective in cooling the kaons. The important slowing down of the
expansion is rather a direct result of a transition through a low pressure
phase which we
believe is the chiral restoration transition. This leads to a softening of
the equation of state and as a result the pressure, which drives the expansion,
is lower than that of the ideal gas. In order to model the effect of chiral
restoration or the softening of the equation of state,  particles in the
fireball should interact via a mean field in addition to the standard
cascade type collisions.
By way of that additional mean field, the equation of state can be softened.
Thus, in our approach the mean field simulates the effect of
chiral restoration.

The model we will use for our calculation is a relativistic transport model
(see e.g. \cite{BKM93}),
which includes nucleon, deltas, pions and kaons. The collisions among these
particles are controlled by known cross sections, where for the nucleon-delta
collisions an improved detailed balance method \cite{DB91} has been used. For
the
kaon nucleon and kaon delta cross section we have assumed a value of $10 \, \rm
 mb$ independent of energy. The $K^+ \pi$ cross section we assume to be
dominated by the $K^*$ intermediate state and we use \cite{Ko81}
\be
\sigma_{K^+\pi} = \frac{\sigma_0}{1 + 4(\sqrt{s} - m_{K^*})^2 / \Gamma^2_{K^*}}
\label{eq.3.1}
\ee
with $\sigma_0 = 60 \, \rm mb$, $m_{K^*} = 895 \, \rm MeV$, and $\Gamma
= 50 \, \rm MeV$.

The mean field is based on the Walecka- ($\sigma - \omega$)  model, with the
possibility to include self interaction terms for the scalar field, if
needed. This model has the advantage, that the necessary field (or
decondensation) energy/pressure associated with a change of the gluon
condensate
 at the chiral restoration transition \cite{KB93b} can be simulated by the
field energy of the scalar sigma-meson at mean field level. It has also been
shown, that the Walecka model shows a very rapid rise in the energy density to
the value of free, massless particles  at a certain critical temperature,
which depends on the parameters used \cite{TGB83}.
Of course the Walecka model is not chirally invariant and one may question
whether it is the correct tool to study the chiral phase transition.
However, since at this point
we are concerned only about the expansion dynamics, a mean field model
which can be parameterized such that the pressure of the system is below that
of an ideal gas serves this purpose. That this is possible
in the Walecka model has been shown in ref. \cite{TGB83}. Incidentally, using
the standard Walecka parameters, the scalar field gives rise to a bag-type
pressure of $\sim 200 \rm \,MeV/fm^3$, which is about the value extracted in
the analysis of lattice calculations \cite{KB93b}.
A more refined
calculation, which addresses questions such as a temperature dependent vector
coupling et c., eventually will have to be done in a chirally symmetric model.

The potential for the kaons is taken of the form
\be
U_K(r) = U_0 \frac{\rho_s}{\rho_0}
\ee
where
the density $\rho_s$ is the scalar density $<\bar{\psi} \psi>$.

Thus, given the energy density as a function of temperature and chemical
potential the Walecka parameters can be adjusted to reproduce that behavior
of the  energy density. For zero chemical potential fairly reliable and model
independent information
for the energy density can be extracted from
lattice gauge calculations \cite{KB93b,KSW91,GKH93}. However, for
HI collisions at presently available energies, the energy density needs to be
known at finite chemical potential. Therefore, lattice results,
which are restricted to vanishing chemical potential can only be used for
guidance and one has to rely on model prediction of the properties of the
chiral
transition at finite chemical potential.
Here, however, we want to follow the opposite, phenomenological approach.
Rather than starting from some
model input,  we will try to extract the properties
of the transition from the available data on kaon production. As demonstrated
in the previous section, it is possible to make statements about the
expansion velocity and hence of the chiral transition independent of the choice
of the potential for the kaons, which also can be fixed from the data.
With both model parameters fixed by data form Si+Au collisions, we then can
make a
prediction for the behavior of the kaon spectra for Au+Au, which have already
been taken and are in the process to be analyzed \cite{E814p}.

In fig. \ref{fig.3.1} we show our result for the kaon production data for
Si+Au.
Together with the data, we also show the prediction of the RQMD model
\cite{SSG89} determined by the E-814 collaboration.
In order to obtain reliable statistics, in our calculation
the kaon spectrum has
been obtained after averaging over rapidity, while the data are taken at
somewhat forward rapidities. A detailed analysis of the rapidity dependence
will
be presented in a more extensive article\footnote{In our schematic model
(section 2) we have checked that in the limit of a Bjorken type fireball, a
soft
component in the kaon spectrum can be obtained as well, provided the
longitudinal
expansion is not too rapid. Since at AGS energies the fireball formed is
somewhere between Bjorken-type and spherical, we a rather confident, that in a
more extensive simulation, the overall picture will not be changed.}.
Depending on whether or not the last data point is to be taken into account
they are reproduced by a potential of $U_0 \simeq 50 \, \rm MeV$. The
parameters
we needed for the Walecka model are
\be
g_v = 5.5, \hspace{1cm} g_s = 9.27
\ee
which is to be compared with the original Walecka parameters which have been
determined from fitting nuclear matter properties
\be
g_v^{orig} = 11.56 \hspace{1cm} g_s^{orig} = 9.27
\ee

Notice that a smaller vector coupling than the original Walecka value
is needed in order to reproduce the kaon data\footnote{There is actually some
freedom in the choice of the couplings. Both $g_s$ and $g_v$ can be changed by
about $25 \%$ without affecting the results, provided that their difference is
kept at about the same value.}.
This is in qualitative agreement with our arguments for the attractive kaon
potential, where also a reduced vector coupling is needed. Notice also, that a
direct comparison with lattice gauge results for the energy density
does not determine the vector coupling since it does not contribute at zero
chemical potential. However, a comparison with measured baryon number
susceptibilities from the lattice also requires a reduction of the vector
coupling around $T_c$ \cite{Kun91},
and a more complete calculation will have to take the
temperature dependence of the vector coupling as given by the lattice gauge
data
into account.
In fig. \ref{fig.3.1} we also show the resulting kaon spectrum obtained without
any mean field for the baryons. Clearly no soft component has developed in this
case.

Finally, our prediction for the kaon spectrum in case of Au+Au. We simulate the
Au+Au collision by assuming a fireball of radius $6.5 \, \rm fm$ and with 350
baryons inside the fireball. In fig. \ref{fig.3.3} we show the
resulting spectrum for this configuration. Again the spectrum develops a
concave shape but at low $m_\bot$ the slope is not as steep as that for the
smaller system, mainly because of the increased rescattering of the
kaons in the bigger fireball.

Before we conclude
we should mention, that in our calculation the protons also show a
soft component at small transverse momentum with  a slope parameter of $T\simeq
80 \, \rm MeV$ which is not seen in experiment. We believe that this is an
artifact of the simplification of our calculation, namely the fact that we have
worked with a constant vector coupling. In a more realistic calculation, the
vector coupling should increase once the system has gone through the chiral
transition giving rise to
additional repulsion in the late stage of the reaction.
( The $g_v = 5.5$ we have used is actually an average between the high
temperature $g_v \simeq 0$ and the low temperature Walecka value
($g_v^{Walecka} \simeq 11$) )
This should give the nucleons
an additional collective push which will eliminate the soft component in the
proton spectra.

\section{Conclusions and Outlook}
In this article we have explained the observed cold component in the
$K^+$ spectrum at AGS energy $Si+Au$ collisions by the E-814 collaboration by
collective cooling. For the cooling mechanism to be successful we had to make
three model assumptions. First, that the kaon feels an
attractive potential in nuclear matter at temperature close to $T_c$. Second,
that the expansion of the fireball created in the collision is slow,
because the
system has to go through the chiral phase transition. And third, that above
$T_c$ the coupling of the vector mesons is reduced. This reduced vector
coupling would then also qualitatively explain the attractive kaon potential.

We have simulated the effect of the chiral transition on the expansion dynamics
with a relativistic mean field model. In order to slow down the system
sufficiently, we had to choose a vector coupling which is smaller than that
extracted from zero temperature properties of nuclear matter.
Having fixed the mean field parameters for the nucleons and kaons by comparing
with the data for $Si+Au$, we have predicted the $K^+$ spectrum for the heavier
system of $Au+Au$. Again the spectrum exhibits a concave shape but with a
larger slope at small transverse mass than that of the small system ($Si +
Au$).
Also the fraction of soft kaons is somewhat
reduced. This is simply a result of the more efficient rescattering in the
bigger system, which knocks more kaons out of the soft component.

If the scenario presented in this article is correct, the smaller system of
Si+Au seems to be better suited to study the effects of chiral symmetry
restoration\footnote{It may also be possible to create a smaller fireball by
triggering on more peripheral Au+Au events.}.
It would, therefore, be desirable if these measurements
could be extended
over a wider range in $p_t$ and rapidity,
such that the connection with the data of the
E-802 collaboration can be made. In particular it would be extremely
interesting to study the kaon spectra as a function of bombarding energy.
Within our scenario we would predict that the
soft component should disappear at
lower bombarding energies, because the chiral symmetry restoration temperature
is not reached and the kaons should not feel an attractive potential.
But also at higher energies, the soft component eventually
should disappear, because for temperatures much larger than $T_c$, when most of
the glue is decondensed, the pressure
increases to that of an ideal gas. As a consequence, the system
expands rapidly and the necessary condition for adiabaticity is not
fulfilled anymore.

Our present calculation will have to be improved in order to include a
dynamical change of the vector coupling, which we have taken to be constant.
In order to ensure energy and
momentum conservation, the mean field model has to be modified. This
modification will be guided by the lattice results on the baryon number
susceptibility. This also will  restrict the choice of mean field parameters
and should cure the problem with the proton spectra.
Furthermore, we are about to investigate, if the measurement of
HBT correlations of the soft kaons could provide independent experimental
information about the long lifetime of the fireball.

In conclusion, we believe that the soft component in the kaon spectra may very
well be the first observed signature for chiral restoration in the laboratory.
The scenario presented in this article may also be applicable to CERN energy
reactions, provided that the temperature reached is not to far above $T_c$.
If this is the case, the analysis of ref. \cite{KB93a} has to be
repeated and it may very well turn out that the observed soft pions at CERN are
yet another result of the chiral restoration transition.

\noindent
{\bf Acknowledgements:} \\
I would like to thank G. Brown for many discussions and
for pointing out, that the pressure is reduced
in a chiral transition. Many discussions about the adiabatic expansion with E.
Shuryak are acknowledged. This work was supported by the US. Dept. of Energy
Grant No. DE-FG02-88ER40388.

\newpage
{\bf Figure captions:}
\begin{figure}[h]
\caption{Initial and final kaon spectra for different expansion times and fixed
kaon potential at $U_0 = - 50 \, \rm MeV$. For orientation the data and RQMD
results of the E-814 group are also shown. The calculated results are all in
arbitrary units and are normalized such that the spectra have the same value at
$m_\bot - m = 0$.}
\label{fig.2.1}
\end{figure}

\begin{figure}[h]
\caption{Initial and final kaon spectra for different kaon potentials and
fixed expansion time $\tau = 10 \, \rm fm/c$. For orientation the data and RQMD
results of the E-814 group are also shown. The calculated results are all in
arbitrary units and are normalized such that the spectra have the same value at
$m_\bot - m = 0$.}
\label{fig.2.2}
\end{figure}

\begin{figure}[h]
\caption{Kaon spectra from transport calculation. Full line is result including
mean fields for Baryons and kaons. The dotted line is the result for kaon mean
field only, no mean field for the baryons
and the dashed line is the pure cascade result. The calculated results are all
in arbitrary units.}
\label{fig.3.1}
\end{figure}

\begin{figure}[h]
\caption{Kaon spectra for Si+Au (full line) and our prediction for Au+Au
(dotted line)}
\label{fig.3.3}
\end{figure}

\end{document}